\documentclass[12pt]{article}
\usepackage[table]{xcolor}
\usepackage{colortbl}
\definecolor{lightgray}{gray}{0.9}
\usepackage{graphicx}
\usepackage{lscape}
\usepackage{citesort}
\usepackage{amssymb}
\usepackage{appendix}
\usepackage{multirow}

\usepackage{graphicx}
\usepackage{lscape}
\usepackage{citesort}
\usepackage{amssymb}
\usepackage{appendix}
\usepackage{multirow}

\usepackage{mathrsfs}
\begin{document}
\def\qq{\langle \bar q q \rangle}
\def\uu{\langle \bar u u \rangle}
\def\dd{\langle \bar d d \rangle}
\def\sp{\langle \bar s s \rangle}
\def\GG{\langle g_s^2 G^2 \rangle}
\def\Tr{\mbox{Tr}}
\def\figt#1#2#3{
        \begin{figure}
        $\left. \right.$
        \vspace*{-2cm}
        \begin{center}
        \includegraphics[width=10cm]{#1}
        \end{center}
        \vspace*{-0.2cm}
        \caption{#3}
        \label{#2}
        \end{figure}
    }

\def\figb#1#2#3{
        \begin{figure}
        $\left. \right.$
        \vspace*{-1cm}
        \begin{center}
        \includegraphics[width=10cm]{#1}
        \end{center}
        \vspace*{-0.2cm}
        \caption{#3}
        \label{#2}
        \end{figure}
                }

\def\ds{\displaystyle}
\def\beq{\begin{equation}}
\def\eeq{\end{equation}}
\def\bea{\begin{eqnarray}}
\def\eea{\end{eqnarray}}
\def\beeq{\begin{eqnarray}}
\def\eeeq{\end{eqnarray}}
\def\ve{\vert}
\def\vel{\left|}
\def\ver{\right|}
\def\nnb{\nonumber}
\def\ga{\left(}
\def\dr{\right)}
\def\aga{\left\{}
\def\adr{\right\}}
\def\lla{\left<}
\def\rra{\right>}
\def\rar{\rightarrow}
\def\lrar{\leftrightarrow}
\def\nnb{\nonumber}
\def\la{\langle}
\def\ra{\rangle}
\def\ba{\begin{array}}
\def\ea{\end{array}}
\def\tr{\mbox{Tr}}
\def\ssp{{\Sigma^{*+}}}
\def\sso{{\Sigma^{*0}}}
\def\ssm{{\Sigma^{*-}}}
\def\xis0{{\Xi^{*0}}}
\def\xism{{\Xi^{*-}}}
\def\qs{\la \bar s s \ra}
\def\qu{\la \bar u u \ra}
\def\qd{\la \bar d d \ra}
\def\qq{\la \bar q q \ra}
\def\gGgG{\la g^2 G^2 \ra}
\def\q{\gamma_5 \not\!q}
\def\x{\gamma_5 \not\!x}
\def\g5{\gamma_5}
\def\sb{S_Q^{cf}}
\def\sd{S_d^{be}}
\def\su{S_u^{ad}}
\def\sbp{{S}_Q^{'cf}}
\def\sdp{{S}_d^{'be}}
\def\sup{{S}_u^{'ad}}
\def\ssp{{S}_s^{'??}}

\def\sig{\sigma_{\mu \nu} \gamma_5 p^\mu q^\nu}
\def\fo{f_0(\frac{s_0}{M^2})}
\def\ffi{f_1(\frac{s_0}{M^2})}
\def\fii{f_2(\frac{s_0}{M^2})}
\def\O{{\cal O}}
\def\sl{{\Sigma^0 \Lambda}}
\def\es{\!\!\! &=& \!\!\!}
\def\ap{\!\!\! &\approx& \!\!\!}
\def\md{\!\!\!\! &\mid& \!\!\!\!}
\def\ar{&+& \!\!\!}
\def\ek{&-& \!\!\!}
\def\kek{\!\!\!&-& \!\!\!}
\def\cp{&\times& \!\!\!}
\def\se{\!\!\! &\simeq& \!\!\!}
\def\eqv{&\equiv& \!\!\!}
\def\kpm{&\pm& \!\!\!}
\def\kmp{&\mp& \!\!\!}
\def\mcdot{\!\cdot\!}
\def\erar{&\rightarrow&}
\def\olra{\stackrel{\leftrightarrow}}
\def\ola{\stackrel{\leftarrow}}
\def\ora{\stackrel{\rightarrow}}

\def\simlt{\stackrel{<}{{}_\sim}}
\def\simgt{\stackrel{>}{{}_\sim}}


\title{
         {\Large
                 {\bf
                     Strong couplings of negative and positive parity nucleons to the heavy baryons and mesons
                 }
         }
      }

\author{\vspace{1cm}\\
{\small K. Azizi$^a$ \thanks {e-mail: kazizi@dogus.edu.tr}\,, Y.
Sarac$^b$
\thanks {e-mail: yasemin.sarac@atilim.edu.tr}\,\,, H.
Sundu$^c$ \thanks {e-mail: hayriye.sundu@kocaeli.edu.tr}} \\
{\small $^a$  Department of Physics, Do\u gu\c s University, Ac{\i}badem-Kad{\i}k\"oy, 34722 Istanbul, Turkey} \\
{\small $^b$ Electrical and Electronics Engineering Department,
Atilim University, 06836 Ankara, Turkey} \\
{\small $^c$ Department of Physics, Kocaeli University, 41380 Izmit,
Turkey}}
\date{}

\begin{titlepage}
\maketitle
\thispagestyle{empty}

\begin{abstract}
The strong coupling form factors related to the strong vertices of the positive and negative parity nucleons with the heavy $\Lambda_{b[c]}[\Sigma_{b[c]}]$ baryons and heavy
 $B^*[D^*]$ vector mesons are calculated using a three-point correlation function. Using the values of the form factors at  $Q^2=-m^2_{meson}$ we compute the strong coupling 
constants among the participating particles.

\end{abstract}

~~~PACS number(s):14.20.Dh, 14.20.Lq, 14.20.Mr, 14.40.Lb, 14.40.Nd, 11.55.Hx
\end{titlepage}

\section{Introduction}

The recently achieved progresses in experimental sector related to
the charm and bottom baryons have provided important clues and
motivations for the theoretical studies on this area. The necessity
for a better understanding of the properties of these baryons such
as their masses, structures and interactions with other particles
have increased the theoretical interests on them. Their various
properties were studied using different methods. For instance their
masses  were studied in Refs.~\cite{Mathur,Ebert,Rosner,Karliner,Wang a3} (see also the references therein) via various methods
such as quenched lattice non-relativistic QCD, the QCD sum rule
approach within the framework of heavy quark effective theory, the
constituent quark model, QCD sum rules and a theoretical approach based
on modeling the color hyperfine interaction. The
Refs.~\cite{Nielsen,Aliev,Diaz,Scholl,Faessler,Cheng,An,Patel,Huang,Wang,chivili,Khodjamirian,Nieves,Gutsche}
and references therein provide some examples in which their strong
and weak decays were studied.

This work provides an analysis of the strong couplings of the heavy
$\Lambda_{b(c)}$ and $\Sigma_{b(c)}$ baryons to the positive parity  nucleon $N$/ negative parity  nucleon $N^*$  and heavy  $B^*$ / $D^*$ vector meson. Here by $N^*$ we mean the 
excited  $N(1535)$ nucleon with $J^P=\frac{1}{2}^-$. Such couplings occur in a low energy regime that preclude us from the usage of the perturbative
approach.  The strong coupling
constants are the basic parameters to  determine the strength of the strong interactions among the participated particles. They
also provide us a better understanding on the structure and nature of the
hadrons participated in the interaction. To improve our
understanding on the perturbative and non-perturbative natures of the
strong interaction they can also provide valuable insights. Furthermore,
these coupling constants may be useful for explanation of the
observation of various exotic events by different collaborations.
Beside these, one may resort to these results in order to explain
the properties of $B^*$ and $D^*$ mesons in nuclear medium. The
nucleon cloud may affect properties of these mesons such as their masses
and decay constants in nuclear medium due to their interactions with nucleons (see for instance the
Refs.~\cite{Hayashigaki,Wang1,Wang2,Kumar,Azizi1,Wang3}). Therefore,
the present study is also helpful to identify the properties of
these particles in nuclear medium.

Here, we calculate the strong form factors defining the strong vertices $\Lambda_bNB^*$, $\Lambda_bN^*B^*$, $\Sigma_bNB^*$, $\Sigma_bN^*B^*$,
$\Lambda_cND^*$,  $\Lambda_cN^*D^*$, $\Sigma_cND^*$ and $\Sigma_cN^*D^*$ in the framework of the QCD sum rule
~\cite{Shifman} as one of the powerful and applicable non-perturbative tools
to hadron physics. By using $Q^2=-m^2_{meson}$,  we then obtain the strong coupling 
constants among the participating particles. This method has been previously applied to  investigate some other vertices (for instance see 
Refs.~\cite{Nielsen,Khodjamirian,Choe,Azizi14,Kazem1} and references therein).

The paper contains three sections. In next section,  we  calculate  the strong
coupling form factors in the context of QCD sum rule approach. Section 3 is
devoted to the numerical analysis of the results and discussion.

\section{The strong coupling form factors}
In this section we calculate the coupling form factors defining the vertices among the hadrons under consideration using the QCD sum rule method. The starting point is to consider 
the following three-point correlation
function:
\begin{eqnarray}\label{CorrelationFunction}
\Pi_{\mu}(q)=i^2 \int d^4x~ \int d^4y~e^{-ip\cdot x}~
e^{ip^{\prime}\cdot y}~{\langle}0| {\cal T}\left (
J_{N}(y)~J_{{\cal M}}^{\mu}(0)~\bar{J}_{{\cal B}}(x)\right)|0{\rangle},
\end{eqnarray}
where ${\cal T}$ is the time ordering operator and  $q=p-p'$ is the transferred momentum. In this equation $J_i$ denote the interpolating fields of different particles, ${\cal M}$
symbolizes the $B^*$ or $D^*$ meson, ${\cal B}$ stands for the
$\Lambda_{b(c)}$ or $\Sigma_{b(c)}$ baryons and $N$ shows the nucleon with both parities. 

The three-point correlation function can be calculated both in terms
of the hadronic degrees of freedom and in terms of the QCD degrees
of freedom. These two different ways of calculations are called as
physical  and OPE sides, respectively. The results obtained from
both sides are equated to acquire the QCD sum rules for the coupling
form factors. For the suppression of the contributions coming from
the higher states and continuum a double Borel transformation with
respect to the variables $p^2$ and $p'^2$ are applied to both sides
of the obtained sum rules.

\subsection{Physical Side}

For the physical side of the calculation one inserts complete sets
of appropriate ${\cal M}$, ${\cal B}$ and $N$ hadronic states, which
have the same quantum numbers as the corresponding interpolating
currents, into the correlation function. Integrals over $x$ and $y$ give
\begin{eqnarray} \label{physicalside}
\Pi_{\mu}^{Phy}(q)&=&\frac{\langle 0 \mid
 J_{N}\mid N(p^{\prime},s^{\prime})\rangle \langle 0 \mid
 J^{\mu}_{{\cal M}}\mid {\cal M}(q)\rangle  \langle N(p^{\prime},s^{\prime}){\cal M}(q)\mid
{\cal B}(p,s)\rangle\langle {\cal B}(p,s)\mid
 \bar{J}_{{\cal B}}\mid 0\rangle }{(p^2-m_{{\cal
B}}^2)(p^{\prime^2}-m_{N}^2)(q^2-m_{{\cal M}}^2)}
 \nonumber \\
&+&\frac{\langle 0 \mid
 J_{N}\mid N^*(p^{\prime},s^{\prime})\rangle \langle 0 \mid
 J^{\mu}_{{\cal M}}\mid {\cal M}(q)\rangle  \langle N^*(p^{\prime},s^{\prime}){\cal M}(q)\mid
{\cal B}(p,s)\rangle\langle {\cal B}(p,s)\mid
 \bar{J}_{{\cal B}}\mid 0\rangle }{(p^2-m_{{\cal
B}}^2)(p^{\prime^2}-m_{N^*}^2)(q^2-m_{{\cal M}}^2)}
 \nonumber \\
&+& \cdots~,
\end{eqnarray}
where $\cdots$ stands for the contributions coming from the higher states and continuum and the contributions of both positive and negative  parity nucleons have been included. 
The matrix elements in this equation are parameterized as
%
\begin{eqnarray}\label{matriselement}
\langle 0 \mid
 J_{N}\mid N(p^{\prime},s^{\prime})\rangle&=&\lambda_N u_N(p^{\prime},s^{\prime}),
\nonumber \\
\langle 0 \mid
 J_{N}\mid N^*(p^{\prime},s^{\prime})\rangle&=&\lambda_{N^*} \gamma_5 u_{N^*}(p^{\prime},s^{\prime}),
 \nonumber \\
\langle {\cal B}_{b(c)}(p,s) \mid
 \bar{J}_{{\cal B}_{b(c)}}\mid
 0\rangle&=&\lambda_{{\cal B}_{b(c)}}\bar{u}_{{\cal B}_{b(c)}}(p, s),
\nonumber \\
\langle 0 \mid
 J^{\mu}_{\cal M}\mid {\cal{M}}(q)\rangle&=&m_{\cal M}f_{\cal M}\epsilon_{\mu}^*,
\nonumber \\
\langle N(p^{\prime},s^{\prime}){\cal M}(q)\mid {\cal
B}(p,s)\rangle&=&\epsilon^{\nu}\bar{u}_N(p^{\prime},s^{\prime})\left[
g_1 \gamma_{\nu}-\frac{i \sigma_{\nu\alpha}}{m_{{\cal B}}+m_N}
q^{\alpha}g_2\right]u_{{\cal B}}(p,s),
\nonumber \\
\langle N^*(p^{\prime},s^{\prime}){\cal M}(q)\mid {\cal
B}(p,s)\rangle&=&\epsilon^{\nu}\bar{u}_{N^*}(p^{\prime},s^{\prime})\gamma_{5}\left[g_1^*\gamma_{\nu}-\frac{i
\sigma_{\nu\alpha}}{m_{{\cal B}}+m_{N^*}}
q^{\alpha}g_2^*\right]u_{{\cal B}}(p,s),
\end{eqnarray}
where  $\lambda_{N(N^*)}$ and $\lambda_{{\cal B}}$ are the
residues of the related baryons, $u_{N(N^*)}$ and $u_{{\cal B}}$ are
the spinors for the nucleon, $\Lambda_b(\Lambda_c)$ and $\Sigma_b
(\Sigma_c)$ baryons; and $f_{{\cal M}}$ represents the leptonic decay
constant of $B^*(D^*)$. Here $g_{1}$ and  $g_{2}$ are strong coupling form factors related to the couplings of the ${\cal B}$ baryon and ${\cal M}$ meson to the positive parity nucleon $N$; and  $g_{1}^*$ and $g_{2}^*$
are those related to the  strong vertices of  ${\cal B}$ baryon and ${\cal M}$ meson with the negative parity nucleon $N^*$. Application of the double Borel
transformation with respect to the  initial and final momenta squared yields
\begin{eqnarray} \label{Borelphysicalside}
\widehat{\textbf{B}}\Pi_{\mu}^{Phy}(q)&=&\lambda_{{\cal
B}}f_{{\cal M}} e^{-\frac{m_{{\cal
B}}^2}{M^2}}~e^{-\frac{m_N^2+m_{N^*}^2}{M^{\prime^2}}}\left[\Phi_1\gamma_{\mu}+\Phi_2\not\!p
q_{\mu}+\Phi_3\not\!q p_{\mu}+\Phi_4 \not\!q
\gamma_{\mu}\right]+\cdots~,
\end{eqnarray}
where 
\begin{eqnarray} \label{Pi1234}
\Phi_1&=&\frac{m_{{\cal M}}}{(m_{{\cal B}}+m_{N^*})(m_{{\cal
M}}^2-q^2)}\Big[e^{\frac{m_N^{*2}}{M^{\prime^2}}}\lambda_N(g_1+g_2)(m_{{\cal
B}}+m_{N^*})\Big(-m_N^2+m_N m_{{\cal B}}+q^2\Big)
\nonumber \\
&+& e^{\frac{m_N^2}{M^{\prime^2}}}\lambda_{N^*}\Big(g_1^*(m_{{\cal
B}}+m_{N^*})+g_2^*(m_{{\cal B}}-m_{N^*})\Big)\Big(
m_{N^*}^2+m_{N^*}m_{{\cal B}}-q^2\Big) \Big],
\nonumber \\
\Phi_2&=&\frac{1}{m_{{\cal M}}(m_{{\cal B}}+m_{N^*})(m_{{\cal
M}}^2-q^2)}\Big[e^{\frac{m_N^{*2}}{M^{\prime^2}}}\lambda_N
\Big(g_1(m_N^2-m_{{\cal B}}^2)+g_2m_{{\cal M}}^2\Big)(m_{{\cal
B}}+m_{N^*})
\nonumber \\
&+&e^{\frac{m_N^2}{M^{\prime^2}}}\lambda_{N^*}(m_{{\cal
B}}-m_{N^*})\Big(g_1^*(m_{{\cal B}}+m_{N^*})^2-g_2^* m_{{\cal
M}}^2\Big) \Big],
\nonumber \\
\Phi_3&=&-\frac{2m_{{\cal M}}}{(m_{{\cal M}}+m_N)(m_{{\cal
B}}+m_{N^*})(m_{{\cal
M}}^2-q^2)}\Big[e^{\frac{m_N^{*2}}{M^{\prime^2}}}\lambda_N(m_{{\cal
B}}+m_{N^*})\Big(g_1(m_{{\cal B}}+m_N)+g_2m_N\Big)
\nonumber \\
&-&e^{\frac{m_N^2}{M^{\prime^2}}}\lambda_{N^*}(m_{{\cal
B}}+m_{N})\Big(g_1^*(m_{{\cal B}}+m_{N^*})-g_2^*m_{N^*}\Big)\Big],
\nonumber \\
\Phi_4&=&\frac{m_{{\cal B}}m_{{\cal M}}}{(m_{{\cal
B}}+m_{N^*})(m_{{\cal
M}}^2-q^2)}\Big[-e^{\frac{m_N^{*2}}{M^{\prime^2}}}\lambda_N(g_1+g_2)(m_{{\cal
B}}+m_{N^*})
\nonumber \\
&+&e^{\frac{m_N^2}{M^{\prime^2}}}\lambda_{N^*}\Big(g_1^*(m_{{\cal
B}}+m_{N^*})+g_2^*(m_{{\cal B}}-m_{N^*})\Big)\Big],
\end{eqnarray}
with $M^2$ and $M^{\prime^2}$
being the Borel mass parameters.

\subsection{OPE Side}
For the OPE side of the calculation, the basic ingredients are the
explicit expressions of the interpolating  currents
in terms of the quark fields, which are taken as
\begin{eqnarray}\label{InterpolatingCurrents}
J_{\Lambda_{b[c]}}(x)&=&\epsilon_{abc}u^{a^T}(x)C\gamma_{5}d^{b}(x)(b[c])^{c}(x),
\nonumber \\
J_{\Sigma_{b[c]}}(x)&=&\epsilon_{abc}\Big(u^{a^T}(x)C\gamma_{\nu}d^{b}(x)\Big)
\gamma_5\gamma_{\nu}(b[c])^{c}(x),
\nonumber \\
J_{N}(y)&=&\varepsilon_{ij\ell}\Big(u^{i^T}(y)C\gamma_{\beta}u^{j}(y)\Big)\gamma_{5}
\gamma_{\beta}d^{\ell}(y),
\nonumber \\
J_{B^*[D^*]}(0)&=&\bar{u}(0)\gamma_{\mu}b[c](0),
\end{eqnarray}
with $C$ being the charge conjugation operator. By replacing these
interpolating currents in  
Eq.~(\ref{CorrelationFunction}) and doing 
contractions of all quark pairs via Wick's theorem, we get
\begin{eqnarray}\label{correlfuncOPELamda}
\Pi_{\mu}^{OPE}(q)&=&i^2\int d^{4}x\int d^{4}ye^{-ip\cdot
x}e^{ip^{\prime}\cdot y}\epsilon_{abc}\epsilon_{ij\ell}
\nonumber \\
&\times&
\Bigg\{\gamma_5\gamma_{\beta}S^{cj}_{d}(y-x)\gamma_{5}CS_{u}^{bi^T}(y-x)C\gamma_{\beta}
S^{ah}_{u}(y) \gamma_{\mu}S_{b[c]}^{h\ell}(-x)
\nonumber \\
&-&\gamma_5\gamma_{\beta}S^{cj}_{d}(y-x)\gamma_{5}CS_{u}^{ai^T}(y-x)C\gamma_{\beta}
S^{bh}_{u}(y) \gamma_{\mu}S_{b[c]}^{h\ell}(-x)
 \Bigg\}~,
\end{eqnarray}
for $\Lambda_bN^{(*)}B^*$ and $\Lambda_cN^{(*)}D^*$ vertices and
\begin{eqnarray}\label{correlfuncOPESigma}
\Pi_{\mu}^{OPE}(q)&=&i^2\int d^{4}x\int d^{4}ye^{-ip\cdot
x}e^{ip^{\prime}\cdot y}\epsilon_{abc}\epsilon_{ij\ell}
\nonumber \\
&\times&
\Bigg\{\gamma_5\gamma_{\beta}S^{cj}_{d}(y-x)\gamma_{\nu}CS_{u}^{bi^T}(y-x)C\gamma_{\beta}
S^{ah}_{u}(y) \gamma_{\mu}S_{b[c]}^{h\ell}(-x)\gamma_{\nu}\gamma_5
\nonumber \\
&-&\gamma_5\gamma_{\beta}S^{cj}_{d}(y-x)\gamma_{\nu}CS_{u}^{ai^T}(y-x)C\gamma_{\beta}
S^{bh}_{u}(y) \gamma_{\mu}S_{b[c]}^{h\ell}(-x)\gamma_{\nu}\gamma_5
 \Bigg\}~,
\end{eqnarray}
for $\Sigma_bN^{(*)}B^*$ and $\Sigma_cN^{(*)}D^*$ vertices.
In these equations, $S^{ij}_{b[c]}(x)$ and $S^{ij}_{u[d]}(x)$ correspond to the heavy
and light quark propagators, respectively. Using the heavy and light quark propagators in coordinate space and after lengthy calculations (for details see Refs.~\cite{Azizi14,Kazem1}), we obtain
\begin{eqnarray}\label{correlfuncOPELast}
\Pi_{\mu}^{OPE}( q)&=&\Pi_1^{OPE}(q^2)\gamma_{\mu}+\Pi_2^{OPE}(q^2)\not\!p
q_{\mu}+\Pi_3^{OPE}(q^2)\not\!q p_{\mu}+\Pi_4^{OPE}(q^2) \not\!q
\gamma_{\mu}\nonumber\\&+&\,\,\mbox{other structures}, 
\end{eqnarray}
where the $\Pi_i(q^2)$ functions contain contributions coming from both the perturbative and
non-perturbative parts and are given as
\begin{eqnarray}\label{QCDside1}
\Pi_i^{OPE}(q^2)=\int^{}_{}ds\int^{}_{}ds^{\prime}
\frac{\rho_i^{pert}(s,s^{\prime},q^2)+\rho_i^{non-pert}(s,s^{\prime},q^2)}{(s-p^2)
(s^{\prime}-p^{\prime^2})}~.
\end{eqnarray}
The spectral densities $\rho_i(s,s',q^2)$ appearing in this equation
are obtained from the imaginary parts of the $\Pi_{i}$ functions as
$\rho_i(s,s',q^2)=\frac{1}{\pi}Im[\Pi_{i}]$. Here, as examples, only the results
of the spectral densities corresponding to the Dirac structure
$\gamma_{\mu}$ for $\Lambda_bNB^*$ vertex are presented, which are
\begin{eqnarray}\label{rho1pert}
\rho_1^{pert}(s,s^{\prime},q^2)&=&
\frac{m_bm_us^{\prime^2}}{64\pi^4{\cal Q}}
\Theta\Big[L_1(s,s^{\prime},q^2)\Big]+\int_{0}^{1}dx
\int_{0}^{1-x}dy \frac{1}{64\pi^4{\cal X}^3}
 \Bigg\{m_b^4x^2({\cal X}^{\prime}+2x)
\nonumber \\
&\times&
 ({\cal
X}+y)+q^4xy\Big[3y({\cal X}^{\prime}-1){\cal X}^{\prime}
 ({\cal X}^{\prime}+4x)-2{\cal X}^{\prime^2}(3x+{\cal
X}^{\prime})+2y^2(15x^2-14x+2 )\Big]
\nonumber \\
&-& 2q^2{\cal X}\Big[xy{\cal
X}^{\prime}\Big(s(2+15x^2-18x)-s^{\prime}({\cal X}^{\prime}-3)
\Big)+ y^2\Big(2sx(1-10x+15x^2)
\nonumber \\
&-& 4sx^2{\cal X}^{\prime^2}-s^{\prime}(1-21x+
41x^2-15x^3)\Big)+s^{\prime}y^3(1-17x+30x^2)  -4sx^2{\cal
X}^{\prime^2}\Big]
\nonumber \\
&+& m_b^3x\Big(m_u(3-5x+2x^2-2xy)+
 3m_d({\cal X}^{\prime}+x){\cal X}\Big)+{\cal
X}^2\Big[s^{\prime^2}y(5y-8x{\cal X}^{\prime}
\nonumber \\
&-&
 34xy-6y^2+15x^2y
+30xy^2)+3s^2x^2(1-4y-6x+5x^2+10xy)+2ss^{\prime}x
\nonumber \\
&\times& (5y-9y^2-4x{\cal X}^{\prime}+
 15x^2y-26xy+ 30xy^2)\Big]+2m_b^2x\Big[q^2\Big(3x{\cal
X}^{\prime}-3y+16xy
\nonumber \\
&+& 8x^2y-4y^2+16xy^2\Big)+{\cal X}\Big(s^{\prime} (3y-3x{\cal
X}^{\prime}-16y+8xy-4y^2 +
16xy^2)
\nonumber \\
&+& 2sx(1-3y-5x+4x^2+8xy)\Big)\Big]+m_b\Big[3m_d {\cal X}\Big(
sx{\cal X}({\cal X}^{\prime}-1)
+y(s^{\prime}u-q^2x)
\nonumber \\
&\times& (3y-1)\Big)+m_u\Big(sx {\cal X}(6-9x + 3x^2-3xy)+y\Big(q^2x(4x
-3x^2+3xy+y-1)
\nonumber \\
&+& 3s^{\prime}{\cal X}{\cal X}^{\prime^2} - 3s^{\prime}y{\cal
X}({\cal X}^{\prime}+2)\Big)\Big)\Big]
\Bigg\}\Theta\Big[L_2(s,s^{\prime},q^2)\Big] ,
\end{eqnarray}
and
\begin{eqnarray}\label{rho1nonpert}
\rho_1^{non-pert}(s,s^{\prime},q^2)&=&\frac{\langle u
\bar{u}\rangle}{16\pi^2{\cal
Q}}\Big[s^{\prime}(m_u-2m_b)-q^2m_d\Big]
\Theta\Big[L_1(s,s^{\prime},q^2)\Big]
\nonumber \\
&+& \int_{0}^{1}dx \int_{0}^{1-x}dy
 \frac{1}{8\pi^2{\cal X}}\Big[\langle d
\bar{d}\rangle\Big(m_d(2x+{\cal X}^{\prime}){\cal X}-m_b(x+{\cal
X}{\cal X}^{\prime})-m_u{\cal X} \Big)
\nonumber \\
&+& \langle u \bar{u}\rangle\Big( m_u(3xy+3x{\cal
X}^{\prime}-y)-m_b(x+{\cal X}^{\prime})-2m_d{\cal X}
\Big)\Big]\Theta\Big[L_2(s,s^{\prime},q^2)\Big]
\nonumber \\
&-& \langle\alpha_s\frac{G^2}{\pi}\rangle\frac{1}{1152\pi^2{\cal
Q}^4}
 \Big[9m_b{\cal
Q}^3(m_d-2m_u) +s^{\prime}{\cal Q}^2
 \Big(3m_b(m_d+m_u)+ 2q^2\Big)
\nonumber \\
&+&
 3s^{\prime^2}\Big(m_b^4
-2q^2m_b(m_b-m_u)+q^4\Big)\Big]
\Theta\Big[L_1(s,s^{\prime},q^2)\Big]
\nonumber \\
&+& \int_{0}^{1}dx \int_{0}^{1-x}dy
\langle\alpha_s\frac{G^2}{\pi}\rangle\frac{1}{192\pi^2{\cal
X}^3}\Big[3{\cal X}^{\prime^3}(2x+{\cal X}^{\prime})+
y^2\Big(15+x(39{\cal X}^{\prime}-20)\Big)
\nonumber \\
&+&  y(2x+{\cal X}^{\prime}) + {\cal X}^{\prime}(11{\cal
X}^{\prime}-1) +6y^3(2x+{\cal X}^{\prime})\Big]
 \Theta\Big[L_2(s,s^{\prime},q^2)\Big]
 \nonumber \\
&-& \frac{1}{192\pi^2{\cal Q}}\Big[m_0^2\langle d
\bar{d}\rangle(6m_b+4m_d)+m_0^2\langle u
\bar{u}\rangle(7m_u-3m_d-18m_b)\Big]
\nonumber \\
&\times&\Theta\Big[L_1(s,s^{\prime},q^2)\Big],
\end{eqnarray}
where
\begin{eqnarray}\label{teta1}
{\cal X}&=&x+y-1,
\nonumber \\
{\cal X}^{\prime}&=&x-1,
\nonumber \\
{\cal Q}&=&m_b^2-q^2,
\nonumber \\
 L_1(s,s^{\prime},q^2)&=&s^{\prime},
\nonumber \\
L_2(s,s^{\prime},q^2)&=&-m_{b}^2x+sx-sx^2+s^{\prime}y+q^2xy-sxy-s^{\prime}xy-s^{\prime}y^2.
\end{eqnarray}
The $\Theta[...]$ in these equations is the unit-step function.
As already stated, the match of the results obtained from physical
and OPE sides of  the correlation function gives the
QCD sum rules for the strong coupling form factors. As examples, for the form factors related to the $\Lambda_bNB^*$ and $\Lambda_bN^{*}B^*$  vertices, we get
\begin{eqnarray}\label{strongcouplingconstants}
g_1(q^2)&=&e^{\frac{m_{\Lambda_b}^2}{M^2}}e^{\frac{m_N^2}{M^{\prime^2}}}\frac{(m_{B^*}^2-q^2)}
{\lambda_N{\cal
H}}\Bigg\{m_{B^*}^2\Big[m_{\Lambda_b}^4(\Pi_3-2\Pi_2)-2{\cal
V}m_Nm_{N^*}\Pi_4 -2m_{\Lambda_b}^3 (m_{N^*}\Pi_2+\Pi_4)
\nonumber \\
&+& m_{\Lambda_b}\Big(2q^2\Pi_4+(m_N^2-m_Nm_{N^*})(m_{N^*}\Pi_3
+2\Pi_4)\Big)
\nonumber \\
&-& m_{\Lambda_b}^2
\Big(m_Nm_{N^*}(2\Pi_2-\Pi_3)+m_{N^*}^2\Pi_3+m_N^2(\Pi_3-2\Pi_2)+2m_{N^*}\Pi_4-2\Pi_1\Big)
\Big]
\nonumber \\
&+&
(m_{\Lambda_b}-m_{N^*})(m_{\Lambda_b}+m_{N^*})\Big[m_{\Lambda_b}m_Nm_{N^*}
{\cal V}\Pi_3-2m_Nm_{N^*}{\cal V}\Pi_4 +
m_{\Lambda_b}^4\Pi_3+2q^2m_{\Lambda_b}\Pi_4
\nonumber \\
&+& m_{\Lambda_b}^2 \Big(2\Pi_1-(m_N{\cal
V}+m_{N^*}^2)\Pi_3\Big)\Big] \Bigg\},
\nonumber \\
g_2(q^2)&=&e^{\frac{m_{\Lambda_b}^2}{M^2}}e^{\frac{m_N^2}{M^{\prime^2}}}\frac{(m_{B^*}^2-q^2)
(m_N+m_{\Lambda_b})}{\lambda_N{\cal
H}}\Bigg\{-m_{\Lambda_b}^5\Pi_3+m_Nm_{\Lambda_b}^4\Pi_3+m_{\Lambda_b}m_{N^*}^3{\cal
V}\Pi_3
\nonumber \\
&+& m_{N^*}m_{\Lambda_b}^3
(2m_{N^*}\Pi_3-m_N\Pi_3-2\Pi_4)+2m_{\Lambda_b}\Big(m_N(m_{N^*}^2
-q^2)+m_{N^*}q^2 \Big)\Pi_4 - 2m_{N^*}^3{\cal
V}\Pi_4
\nonumber \\
&+& m_{B^*}^2(m_{\Lambda_b}-{\cal V})\Big(m_{\Lambda_b}
(2m_{\Lambda_b}\Pi_2-m_{\Lambda_b}\Pi_3 +m_{N^*}\Pi_3) +
2(m_{\Lambda_b}-m_{N^*})\Pi_4\Big)
\nonumber \\
&-& m_{\Lambda_b}^2\Big(m_{N^*}(m_Nm_{N^*}\Pi_3-
4m_N\Pi_4+4m_{N^*}\Pi_4)+2{\cal V}\Pi_1\Big)
  \Bigg\},
  \nonumber \\
g_1^*(q^2)&=&e^{\frac{m_{\Lambda_b}^2}{M^2}}e^{\frac{m_N^2}{M^{\prime^2}}}\frac{(m_{B^*}^2-q^2)}{
\lambda_{N^*}{\cal
H}}\Bigg\{(m_{\Lambda_b}-m_N)(m_{\Lambda_b}+m_N)\Big[m_{\Lambda_b}^4\Pi_3
+m_{\Lambda_b}m_Nm_{N^*}{\cal V} \Pi_3
\nonumber \\
&+& 2m_Nm_{N^*}{\cal
V}\Pi_4+2m_{\Lambda_b}q^2\Pi_4+m_{\Lambda_b}^2\Big(
2\Pi_1-(m_N{\cal V}+m_{N^*}^2)\Pi_3\Big)\Big]
\nonumber \\
&+& m_{B^*}^2\Big[m_Nm_{N^*}m_{\Lambda_b}{\cal V}
\Pi_3+2m_{\Lambda_b}^3(m_N\Pi_2-\Pi_4)-2m_Nm_{N^*}{\cal V} \Pi_4
\nonumber \\
&+& m_{\Lambda_b}^4(\Pi_3-2\Pi_2)-2m_{\Lambda_b}\Big(m_{N^*}{\cal
V}-q^2\Big)\Pi_4+m_{\Lambda_b}^2\Big(m_{N^*}^2
(2\Pi_2-\Pi_3)-m_N^2\Pi_3
\nonumber \\
&+& m_N(2\Pi_4+m_{N^*}\Pi_3-2m_{N^*}\Pi_2)+2\Pi_1\Big)\Big]
\Bigg\},
\nonumber \\
g_2^*(q^2)&=&e^{\frac{m_{\Lambda_b}^2}{M^2}}e^{\frac{m_N^2}{M^{\prime^2}}}\frac{(m_{B^*}^2-q^2)
(m_{\Lambda_b}+m_{N^*})}{ \lambda_{N^*}{\cal
H}}\Bigg\{\Big(m_{\Lambda_b}^5+m_{\Lambda_b}m_N^3{\cal
V}+m_{\Lambda_b}^4m_{N^*}\Big)\Pi_3+m_{\Lambda_b}^3m_N
\nonumber \\
&\times& \Big(({\cal V}-m_N)\Pi_3-2\Pi_4\Big)-2m_N^3{\cal
V}\Pi_4+2m_{\Lambda_b}\Big(m_N^2m_{N^*}+q^2{\cal
V}\Big)\Pi_4-m_{B^*}^2(m_{\Lambda_b}-{\cal V})
\nonumber \\
&\times&
\Big[m_{\Lambda_b}\Big(2m_{\Lambda_b}\Pi_2-(m_{\Lambda_b}+m_N)\Pi_3\Big)+2(m_{\Lambda_b}+m_N)
\Pi_4\Big]-m_{\Lambda_b}^2\Big[m_N\Big(m_Nm_{N^*}\Pi_3
\nonumber \\
&-& 4m_N\Pi_4+4m_{N^*}\Pi_4\Big)-2{\cal V}\Pi_1\Big] \Bigg\},
\end{eqnarray}
where
\begin{eqnarray}\label{strongcouplingconstants}
{\cal
H}&=&2f_{B^*}\lambda_{\Lambda_b}m_{B^*}m_{\Lambda_b}^2(m_{\Lambda_b}-{\cal
V})(m_N+m_{N^*})
\Big(m_{B^*}^2+m_{\Lambda_b}^2-m_Nm_{\Lambda_b}-m_N^2
\nonumber \\
&+& m_{N^*}(m_N+m_{\Lambda_b})-m_{N^*}^2\Big),
\nonumber \\
 {\cal V}&=&(m_N-m_{N^*}).
\end{eqnarray}

\section{Numerical results}

 To numerically analyze the sum rules for the strong coupling form
factors and to find their behavior  with respect to  $Q^2=-q^2$ we need some inputs as presented in table 1.
\begin{table}[ht]\label{Table1}
\centering \rowcolors{1}{lightgray}{white}
\begin{tabular}{cc}
\hline \hline
   Parameters  &  Values
           \\
\hline \hline $  \langle \bar{u}u\rangle(1~GeV)=\langle
\bar{d}d\rangle(1~GeV)$& $-(0.24\pm0.01)^3 $ $\mbox{GeV$^3$}$
 \cite{Ioffe} \\
$ \langle\frac{\alpha_sG^2}{\pi}\rangle $       &
$(0.012\pm0.004)$ $~\mbox{GeV$^4$}$
\cite{belyaev}   \\
$ m_0^2(1~GeV) $       & $(0.8\pm0.2)$ $~\mbox{GeV$^2$}$
\cite{belyaev}   \\
$m_{b}$              & $(4.18\pm0.03)~\mbox{GeV}$\cite{Olive}\\
$m_{c}$              & $(1.275\pm0.025)~\mbox{GeV}$\cite{Olive}\\
$m_{d}$              & $4.8^{+0.5}_{-0.3}~\mbox{MeV}$\cite{Olive}\\
$ m_{u} $            &$2.3^{+0.7}_{-0.5}~\mbox{MeV}$ \cite{Olive}\\
$ m_{B^*}$    &   $ (5325.2\pm0.4)~\mbox{MeV}$ \cite{Olive}  \\
$ m_{D^*}$    &   $ (2006.96\pm0.10)~\mbox{MeV}$ \cite{Olive}  \\
$ m_{N} $      &   $ (938.272046\pm0.000021)~\mbox{MeV}$  \cite{Olive} \\
$ m_{N^*} $      &   $ 1525~TO~1535~\mbox{MeV}$  \cite{Olive} \\
$ m_{\Lambda_b} $      &   $ (5619.5\pm0.4) ~\mbox{MeV} $ \cite{Olive}  \\
$ m_{\Lambda_c} $      &   $ (2286.46\pm0.14) ~\mbox{MeV} $ \cite{Olive}  \\
$ m_{\Sigma_b} $      &   $ (5811.3\pm1.9) ~\mbox{MeV} $ \cite{Olive}  \\
$ m_{\Sigma_c} $      &   $(2452.9\pm 0.4) ~\mbox{MeV} $ \cite{Olive}  \\
$ f_{B^*} $      &   $(210.3^{+0.1}_{-1.8}) ~\mbox{MeV}$
\cite{Khodjamirian1} \\
$ f_{D^*} $      &   $(241.9^{+10.1}_{-12.1}) ~\mbox{MeV}$  \cite{Khodjamirian1} \\
$ \lambda_{N}^2 $      &   $0.0011\pm0.0005  ~\mbox{GeV$^6$}$  \cite{Azizi2} \\
$\lambda_{N^*} $  &   $0.019\pm0.0006  ~\mbox{GeV$^3$}$  \cite{Azizi3} \\
$ \lambda_{\Lambda_b} $      &   $(3.85\pm0.56)10^{-2}$ $\mbox{GeV$^3$}$  \cite{Azizi} \\
$ \lambda_{\Sigma_b} $      &   $(0.062\pm0.018)$ $\mbox{GeV$^3$}$  \cite{ZGWang} \\
$ \lambda_{\Lambda_c} $      &   $(3.34\pm0.47)10^{-2}$ $\mbox{GeV$^3$}$  \cite{Azizi} \\
$ \lambda_{\Sigma_c} $      &   $(0.045\pm0.015)$ $\mbox{GeV$^3$}$  \cite{ZGWang} \\
 \hline \hline
\end{tabular}
\caption{Input parameters used in the calculations.}
\end{table}
Besides,  we
also need to determine the working regions corresponding to the four
auxiliary parameters, $M^2$, $M'^2$, $s_0$ and $s'_0$. The  $M^2$ and
$M'^2$ emerge from the double Borel transformation and $s_0$ and
$s'_0$ originate from continuum subtraction. These are auxiliary
parameters, therefore,  we need a region of them through which the
strong coupling form factors have weak dependency on these parameters. The
continuum thresholds are in relation
with the first excited states in the initial and final channels. To
determine them the energy that characterizes the beginning of the
continuum is considered. Table 2 presents
intervals of the continuum thresholds used in the calculations. To determine the working
regions for the Borel mass parameters, we need to take into account
the criteria that the contributions of the higher states and
continuum are sufficiently suppressed and the contributions of the
operators with higher dimensions are small. The intervals obtained
based on these considerations are also given in table 2.
\begin{table}[h]
\renewcommand{\arraystretch}{1.5}
\addtolength{\arraycolsep}{3pt}
$$
\begin{array}{|c|c|c|c|c|}
\hline \hline
       \mbox{Vertex}    &s_0(GeV^2) & s_0^{\prime}(GeV^2) &M^2 (GeV^2)  &  M^{\prime^2}(GeV^2)     \\
\hline
  \mbox{$\Lambda_bN^{(*)}B^*$} &32.71\leq s_0 \leq 35.04&1.04\leq s_0^{\prime}\leq
  1.99&10\leq M^2\leq20&1\leq M^{\prime^2}\leq 3 \\
  \hline
  \mbox{$\Sigma_bN^{(*)}B^*$} &34.91\leq s_0 \leq 37.40&1.04\leq s_0^{\prime}\leq
  1.99&10\leq M^2\leq20&1\leq M^{\prime^2}\leq 3 \\
  \hline
   \mbox{$\Lambda_cN^{(*)}D^*$} &5.71\leq s_0 \leq 6.72&1.04\leq s_0^{\prime}\leq
  1.99&2\leq M^2\leq6&1\leq M^{\prime^2}\leq 3 \\
  \hline
  \mbox{$\Sigma_cN^{(*)}D^*$} &6.51\leq s_0 \leq 7.62&1.04\leq s_0^{\prime}\leq
  1.99&2\leq M^2\leq6&1\leq M^{\prime^2}\leq 3 \\
                        \hline \hline
\end{array}
$$
\caption{Working intervals for auxiliary parameters.}
\label{workingregions}
\renewcommand{\arraystretch}{1}
\addtolength{\arraycolsep}{-1.0pt}
\end{table}

The determination of the working regions of auxiliary parameters is
followed by the usage of them together with the other input
parameters to obtain the variation of the coupling form factors as a
function of $Q^2$. For this purpose, the following fit function is
applied
\begin{eqnarray}\label{fitfunc}
g_{{\cal B}N{\cal
M}}(Q^2)=c_1+c_2\exp\Big[-\frac{Q^2}{c_3}\Big].
\end{eqnarray}
where $c_1$, $c_2$ and $c_3$ for different vertices are given in tables \ref{fitparam}-\ref{fitparam3}.
This fit function is used to attain the coupling constants at
$Q^2=-m_{{\cal M}}^2$ for all structures. The results  for coupling constants are
presented in table \ref{couplingconstant}. The presented errors in
the results arise due to the uncertainties of the input
parameters as well as uncertainties coming from the determination of
the working regions of the auxiliary parameters. From this table we see that the maximum value belongs to the coupling constant $g_2^*$ associated to the vertex $\Lambda_bN^*B^*$  and the minimum
value corresponds to the coupling  $g_1$ related to the $\Lambda_cND^*$ vertex.
\begin{table}[h]
\renewcommand{\arraystretch}{1.5}
\addtolength{\arraycolsep}{3pt}
$$
\begin{array}{|c|c|c|c|c|c|}
\hline \hline
       \mbox{}    & c_1 & c_2 &c_3(\mbox{GeV$^2$})      \\
\hline
  \mbox{$g_1(Q^2)$} &-2.44\pm 0.68&-0.34\pm0.10&-17.88\pm5.18 \\
  \hline
  \mbox{$g_2(Q^2)$} &22.92\pm 6.64&3.87\pm1.12&16.85\pm4.89 \\
  \hline
   \mbox{$g_1^*(Q^2)$} &-6.21\pm 1.73&-26.76\pm8.01&-193.72\pm56.17 \\
  \hline
  \mbox{$g_2^*(Q^2)$} &88.27\pm 25.60&9.65\pm2.70&24.32\pm7.05\\
                        \hline \hline
\end{array}
$$
\caption{Parameters appearing in the fit function of the coupling
form factor related to the  $\Lambda_bN^{(*)}B^*$ vertex.} \label{fitparam}
\renewcommand{\arraystretch}{1}
\addtolength{\arraycolsep}{-1.0pt}
\end{table}
\begin{table}[h]
\renewcommand{\arraystretch}{1.5}
\addtolength{\arraycolsep}{3pt}
$$
\begin{array}{|c|c|c|c|c|c|}
\hline \hline
       \mbox{}    & c_1 & c_2 &c_3(\mbox{GeV$^2$})      \\
\hline
  \mbox{$g_1(Q^2)$} &297.08\pm 89.12&-282.66\pm81.97&-225.11\pm67.53 \\
  \hline
  \mbox{$g_2(Q^2)$} &-18.12\pm 5.07&-3.82\pm1.14&14.06\pm4.08 \\
  \hline
   \mbox{$g_1^*(Q^2)$} &87.60\pm 25.40&-82.34\pm23.87&-24.88\pm7.21 \\
  \hline
  \mbox{$g_2^*(Q^2)$} &31.80\pm 9.22&0.90\pm0.26&-6.70\pm1.94\\
                        \hline \hline
\end{array}
$$
\caption{Parameters appearing in the fit function of the coupling
form factor  related to the  $\Sigma_bN^{(*)}B^*$ vertex.} \label{fitparam1}
\renewcommand{\arraystretch}{1pt}
\addtolength{\arraycolsep}{1.0pt}
\end{table}
\begin{table}[h]
\renewcommand{\arraystretch}{1.5}
\addtolength{\arraycolsep}{3pt}
$$
\begin{array}{|c|c|c|c|c|c|}
\hline \hline
       \mbox{}    & c_1 & c_2 &c_3(\mbox{GeV$^2$})     \\
\hline
  \mbox{$g_1(Q^2)$} &1.28\pm 0.36&0.92\pm0.27&-155.98\pm45.24 \\
  \hline
  \mbox{$g_2(Q^2)$} &3.88\pm 1.13&1.27\pm0.38&3.60\pm1.04 \\
  \hline
   \mbox{$g_1^*(Q^2)$} &3.01\pm 0.87&(17.97\pm5.21)10^{-4}&-2.77\pm0.80 \\
  \hline
  \mbox{$g_2^*(Q^2)$} &11.52\pm 3.23&-2.38\pm0.71&2.26\pm0.66\\
                        \hline \hline
\end{array}
$$
\caption{Parameters appearing in the fit function of the coupling
form factor  related to the  $\Lambda_cN^{(*)}D^*$ vertex.} \label{fitparam2}
\renewcommand{\arraystretch}{1}
\addtolength{\arraycolsep}{-1.0pt}
\end{table}
\begin{table}[h]
\renewcommand{\arraystretch}{1.5}
\addtolength{\arraycolsep}{3pt}
$$
\begin{array}{|c|c|c|c|c|c|}
\hline \hline
       \mbox{}    & c_1 & c_2 &c_3(\mbox{GeV$^2$})    \\
\hline
  \mbox{$g_1(Q^2)$} &-6.94\pm 2.01&11.18\pm3.35&8.30\pm2.41 \\
  \hline
  \mbox{$g_2(Q^2)$} &-4.64\pm 1.35&-(1.41\pm0.4)10^{-2}&1.53\pm0.45 \\
  \hline
   \mbox{$g_1^*(Q^2)$} &26.37\pm 7.65&-23.18\pm6.49&-11.03\pm3.20 \\
  \hline
  \mbox{$g_2^*(Q^2)$} &15.47\pm 4.48&2.22\pm0.67&-6.34\pm1.84\\
                        \hline \hline
\end{array}
$$
\caption{Parameters appearing in the fit function of the coupling
form factor  related to the  $\Sigma_cN^{(*)}D^*$ vertex.} \label{fitparam3}
\renewcommand{\arraystretch}{1}
\addtolength{\arraycolsep}{-1.0pt}
\end{table}
\begin{table}[h]
\renewcommand{\arraystretch}{1.5}
\addtolength{\arraycolsep}{3pt}
$$
\begin{array}{|c|c|c|c|c|c|}
\hline \hline
     \mbox{Vertex}     & |g_1(Q^2=-m_{{\cal M}}^2)|&  |g_2 (Q^2=-m_{{\cal M}}^2)| &
     |g_1^*(Q^2=-m_{{\cal M}}^2)|&  |g_2^* (Q^2=-m_{{\cal M}}^2)|   \\
\hline
  \mbox{$\Lambda_bN^{(*)}B^*$} &2.51\pm0.75&43.73\pm13.11&29.32\pm8.78&119.26\pm35.77 \\
  \hline
  \mbox{$\Sigma_bN^{(*)}B^*$} &47.87\pm14.35&46.83\pm14.04&61.25\pm18.37&31.81\pm9.54 \\
   \hline
  \mbox{$\Lambda_cN^{(*)}D^*$} &2.05\pm0.61&7.78\pm2.33&3.01\pm0.90&2.60\pm0.78 \\
  \hline
  \mbox{$\Sigma_cN^{(*)}D^*$}  &11.21\pm3.36&4.64\pm 1.39&10.29\pm3.08&16.65\pm4.99\\
                         \hline \hline
\end{array}
$$
\caption{Values of the strong coupling constants for the vertices under consideration.}
\label{couplingconstant}
\renewcommand{\arraystretch}{1}
\addtolength{\arraycolsep}{-1.0pt}
\end{table}

In conclusion, we calculated the strong  coupling constants related to the vertices $\Lambda_bNB^*$, $\Lambda_bN^*B^*$, $\Sigma_bNB^*$, $\Sigma_bN^*B^*$,
$\Lambda_cND^*$,  $\Lambda_cN^*D^*$, $\Sigma_cND^*$ and $\Sigma_cN^*D^*$  in the framework  QCD sum rules. Our results may be checked via other non-perturbative approaches. 
The presented results can be helpful to
explain different exotic events observed via different experiments. These results may also be useful in the
analysis of the results of  heavy ion collision experiments as well as  in exact
determinations of the modifications in the masses, decay constants
and other parameters of the $B^*$ and $D^*$ mesons in nuclear
medium.

\section{Acknowledgment}
This work has been supported in part by the Scientific and Technological
Research Council of Turkey (TUBITAK) under the grant no: 114F018.

\end{document}